\pdfoutput=1
\documentclass[ reprint,twocolumn,groupedaddress]{revtex4-2}
\usepackage[T1]{fontenc}
\usepackage{amsmath}
\usepackage{amssymb}
\usepackage{graphicx}
\usepackage{dcolumn}
\usepackage{bm}

\begin{document}

\title{Scatter-Free Perfect Reflection in Multichannel Monolithic High-Contrast Gratings: Direct--Indirect Fourier-Harmonic Interference and Dimensionality of the Solution Set}

\author{Sun-Goo Lee}
\email{sungooleee@gmail.com}
\affiliation{Department of Data Information and Physics, Kongju National University, Gongju, 32588, South Korea}
\affiliation{Institute of Application and Fusion for Light, Kongju National University, Cheonan, 31080, South Korea}

\author{Wook-Jae Lee}
\email{wookjaelee@kongju.ac.kr}
\affiliation{Department of Data Information and Physics, Kongju National University, Gongju, 32588, South Korea}
\affiliation{Institute of Application and Fusion for Light, Kongju National University, Cheonan, 31080, South Korea}
\date{\today}

\begin{abstract}
Monolithic high-contrast gratings (MHCGs) can exhibit scatter-free perfect reflection despite open higher-order substrate diffraction channels, but the underlying mechanism has not been fully clarified. Using a Fourier-harmonic description, we show that all open higher-order channels in the high-index substrate can be simultaneously extinguished through destructive interference between the radiation fields generated by the direct and indirect source contributions. By contrast, the zeroth-order transmission vanishes when the modulation-induced zeroth-order field interferes destructively with the background transmitted field of the incident wave. Together, these cancellations yield unit zeroth-order reflectance in lossless MHCGs. The near invariance of the internal field harmonics accounts for the remarkable stability of these conditions against changes in the substrate refractive index. We further show that the number of symmetry-compatible open channels determines the generic codimension of the perfect-reflection solution set. Depending on the number of independent design parameters relative to this codimension, the solutions form continuous curves, appear as isolated points, or are generically absent. Introducing an additional geometrical parameter restores isolated perfect-reflection solutions when an additional diffraction channel opens. The same cancellation mechanisms and channel counting apply to both TE and TM polarizations.
\end{abstract}


\maketitle

\section{Introduction}

High-contrast gratings (HCGs) have attracted considerable attention as versatile optical platforms capable of producing a wide variety of spectral responses~\cite{Mateus2004,Huang2007,ChangHasnain2012,Qiao2018}. These responses originate from resonant coupling between externally incident light and eigenmodes confined within the grating layer~\cite{Karagodsky2010,Karagodsky2011,Karagodsky2012}. Monolithic HCGs (MHCGs) constitute a distinct HCG configuration, in which the grating structure is formed by directly etching a high-index substrate~\cite{Gebski2015,Czyszanowski2017}. The grating layer is therefore no longer bounded by low-index media on both sides, which provide the vertical confinement of the eigenmodes in conventional HCGs. Moreover, the high-index substrate can support multiple propagating diffraction channels. Nevertheless, MHCGs can exhibit near-unity zeroth-order reflection back into the substrate, with essentially no power scattered into the open higher-order diffraction channels. This distinctive reflection behavior has made MHCGs promising compact alternatives to conventional multilayer dielectric mirrors, particularly distributed Bragg reflectors (DBRs) used in vertical-cavity surface-emitting lasers (VCSELs)~\cite{Gebski2019,Hong2021,Heermeier2025LPR,Heermeier2025Optica}. MHCGs have since been extended to planar microcavities~\cite{Czyszanowski2020}, polariton lasers~\cite{Kim2019}, planar focusing reflectors~\cite{Komar2020,Komar2021AMI}, and metal-integrated electrodes with high infrared transmission~\cite{Ekielski2024,Bogdanowicz2026}.

Scatter-free perfect-reflection states in MHCGs can be found by optimizing the geometrical parameters~\cite{Marciniak2016}. Previous studies have explained the absence of scattering in terms of the grating impedance or admittance matrix~\cite{Dems2017} and the nearly flat superposition of grating modes at the input interface~\cite{Pruszynska2023}. These descriptions account for the overall reflection behavior, but do not explicitly show how the distinct coupling contributions within the grating cancel to extinguish each open higher-order diffraction channel. A complementary perspective is provided by Fourier-component engineering in periodic photonic lattices, where individual Fourier components of the dielectric modulation couple field harmonics with definite in-plane momentum differences, allowing the contributions to a given diffraction order to be distinguished by the coupling pathways involved~\cite{SGLee2021PRL,SGLee2021Nanoph,SGLee2021PR}. For radiation from resonant Bloch modes, these contributions have been further separated into direct and indirect terms, whose destructive interference can suppress radiation into a channel even when the corresponding field harmonic remains finite~\cite{SGLee2026Nano}. This provides a basis for examining how each open diffraction channel is extinguished in the scattering of an externally incident wave by an MHCG.

\begin{figure*}[t]
\centering
\includegraphics[width=\textwidth]{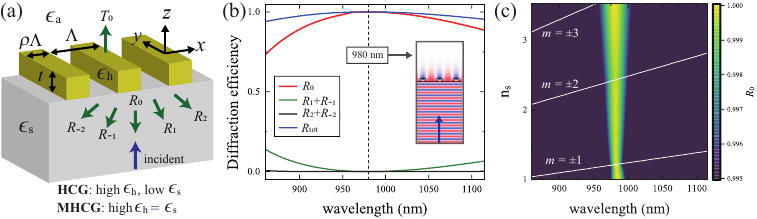}
\caption{(a) Schematic of the one-dimensional HCG and the propagating diffraction channels for normal incidence from the substrate. The illustrated channels correspond to the MHCG at $980$~nm, where the air side supports only the zeroth transmitted order, whereas the substrate supports the reflected orders $m=0,\pm1,\pm2$. (b), (c) RCWA results for the TE-optimized MHCG with the structural parameters reported in Refs.~\cite{Gebski2015,Dems2017}. (b) Diffraction efficiencies of the reflected orders as a function of wavelength. The dashed line marks $980$~nm, where the higher-order reflection vanishes and $R_0$ reaches unity. $R_{\mathrm{tot}}$ denotes the total reflected efficiency. The inset shows the real part of $E_y$ at $980$~nm obtained by the FEM simulation for the incident wave entering from the substrate. (c) Zeroth-order efficiency $R_0$ as a function of wavelength and substrate refractive index $n_s$ for the same structure, showing the remarkable stability of the high-reflectivity state against substrate variations. The white lines mark the Rayleigh thresholds $n_s=|m|\lambda/\Lambda$ of the substrate orders $m=\pm1$, $\pm2$, and $\pm3$.}
\label{fig1}
\end{figure*}

In this study, we elucidate the mechanism of scatter-free perfect reflection in MHCGs. Starting from the coupled equations for the field harmonics, we formulate the radiative source term for each diffraction order in terms of the Fourier components of the dielectric modulation and the field harmonics coupled through them. We separate this source into a direct contribution from the zeroth field harmonic and indirect contributions from the remaining field harmonics. We show that the radiation fields generated by these contributions destructively interfere, simultaneously extinguishing all open higher-order diffraction channels even though the corresponding field harmonics remain finite within the grating layer. In contrast, the zeroth-order transmission vanishes through destructive interference between the modulation-induced zeroth-order field and the background transmitted field of the incident wave. The simultaneous satisfaction of these channel-specific cancellation conditions leaves only the zeroth-order reflected wave. We further examine the dependence of these conditions on the substrate refractive index and show that their remarkable stability originates from the near invariance of the field harmonics at the scatter-free state. We also establish how the number of symmetry-compatible open channels determines the generic codimension of the perfect-reflection solution set. Together with the number of independent design parameters, this counting explains why the solutions form continuous curves, appear as isolated points, or are generically absent. We verify these predictions numerically and demonstrate that an additional geometrical parameter restores isolated perfect-reflection solutions when an additional substrate diffraction channel opens. Both the cancellation mechanisms and the channel counting are demonstrated for TE and TM polarizations.

\section{Grating structure and the physical perspective}

Figure~\ref{fig1}(a) shows the geometry of a one-dimensional HCG. The grating layer has a thickness $t$ and a period $\Lambda$ and is placed between air with dielectric constant $\epsilon_a=1$ and a homogeneous substrate with dielectric constant $\epsilon_s$. The dielectric profile is periodic along the $x$ direction and uniform along the $y$ direction. Within each period, a region of width $\rho\Lambda$ has the higher dielectric constant $\epsilon_h$, while the remaining region is air, where $\rho$ is the duty cycle. The grating corresponds to a monolithic configuration when the region with dielectric constant $\epsilon_h$ and the substrate are formed from the same material, such that $\epsilon_h=\epsilon_s$. A transverse-electric (TE) polarized plane wave, with its electric field along the $y$ direction, is normally incident from the substrate. The transverse-magnetic (TM) polarization, with its magnetic field along the $y$ direction, is discussed later in this study.

Conventional HCGs are operated in the zeroth-order diffraction regime, so that only the zeroth diffraction order is propagating on either side of the grating. By contrast, an MHCG can operate with only the zeroth diffraction order propagating in air, while the high-dielectric-constant substrate supports multiple propagating diffraction orders. Nevertheless, perfect zeroth-order reflection can be obtained with no power scattered into the open higher-order diffraction channels. To illustrate this unusual reflection behavior, Figs.~\ref{fig1}(b) and \ref{fig1}(c) reproduce the characteristic responses of the TE-optimized MHCG reported in Refs.~\cite{Gebski2015,Dems2017}, calculated by rigorous coupled-wave analysis (RCWA). The grating period, duty cycle, and thickness of this structure are $\Lambda=817$~nm, $\rho=0.350$, and $t=164$~nm, respectively, and the dielectric constants are $\epsilon_h=\epsilon_s=12.3904$ and $\epsilon_a=1$, taken as lossless and nondispersive. Figure~\ref{fig1}(b) shows the diffraction efficiencies of the reflected orders as a function of wavelength. At the wavelength $\lambda=980$~nm, the higher-order efficiencies vanish and the zeroth-order efficiency reaches unity, yielding a scatter-free reflected plane wave despite the presence of open higher-order channels in the substrate. This is seen directly in the inset, which shows the field distribution at $980$~nm obtained by the finite-element method (FEM) simulation. Away from the grating, the substrate contains only the incident and the zeroth-order reflected waves, which form a standing wave uniform along $x$, and no propagating power is transmitted into the air. Figure~\ref{fig1}(c) shows the zeroth-order efficiency $R_0$ as a function of wavelength and substrate refractive index $n_s=\sqrt{\epsilon_s}$, following the notation of Ref.~\cite{Dems2017}. The high-reflectivity state remains remarkably stable in both wavelength and reflectance as $n_s$ varies from unity to high-index values, even as higher-order diffraction channels cross their cutoff conditions.

The diffraction response can be calculated using full-wave methods or grating-specific formulations~\cite{Jin2014,Taflove1995,Moharam1981,Moharam1995,Petit1980}. Here, RCWA provides the complex amplitudes of the reflected and transmitted orders, from which the efficiencies in Figs.~\ref{fig1}(b) and \ref{fig1}(c) are obtained. These amplitudes establish which channels are extinguished, but do not separately identify the coupling contributions within the grating that produce their extinction. The absence of scattering in MHCGs was previously analyzed by matching the fields at the two interfaces of the grating layer and expressing the reflection problem in terms of an impedance matrix for TM polarization and an admittance matrix for TE polarization~\cite{Dems2017}. In this formulation, the grating-dependent matrix is independent of the substrate refractive index, while the substrate dependence enters separately through the diffraction-wave impedances or admittances. At the scatter-free high-reflectivity state, the particular structure of these matrices gives a zeroth-order reflectance that remains nearly unity as the substrate refractive index is varied. For a lossless grating, this leaves essentially no power in the other propagating channels. The same behavior was also interpreted in terms of the grating modes at the reflection interface, whose superposition forms a nearly flat field profile that couples predominantly to the zeroth-order reflected plane wave and only weakly to the higher diffraction orders~\cite{Pruszynska2023}.

These descriptions characterize the reflection through the boundary response of the grating or the superposition of its modes. Our aim is to resolve the coupling contributions that generate the outgoing amplitude in each diffraction channel. We therefore express the field inside the grating in spatial harmonics and identify the source driving each harmonic through the Fourier components of the dielectric modulation. A field harmonic inside the grating is coupled to other harmonics, whereas the corresponding diffraction order propagates or decays independently in each homogeneous exterior region. Consequently, the extinction of an outgoing diffraction channel does not require the associated field harmonic to vanish throughout the grating. In the following section, we relate the coupled-harmonic source terms to the outgoing diffraction amplitudes, providing the basis for examining this extinction at the amplitude level.

\section{Fourier-harmonic description of the diffraction channels}
\label{sec:fourier}

To analyze the reflection and transmission properties of the grating, we start from the wave equation governing the TE field,
\begin{equation}
\left(\frac{\partial^2}{\partial x^2}+\frac{\partial^2}{\partial z^2}\right)E_y(x,z)+k_0^2\epsilon(x,z)E_y(x,z)=0,
\label{eq:wave}
\end{equation}
where $E_y(x,z)$ is the electric-field component along the $y$ direction and $k_0$ is the free-space wavenumber. Within the grating layer, the dielectric function is expanded as
\begin{equation}
\epsilon(x,z)=\sum_{p=-\infty}^{\infty}\gamma_p(z)e^{ipKx},
\label{eq:eps}
\end{equation}
where $K=2\pi/\Lambda$, $\gamma_0$ is the spatially averaged dielectric constant, and $\gamma_p$ for $p\neq0$ denotes the $p$th Fourier component of the dielectric modulation. For the centered binary grating considered here, the region with dielectric constant $\epsilon_h$ occupies a fraction $\rho$ of the period, while the remaining region is air. The Fourier coefficients are real and even and are given by $\gamma_0=\epsilon_a+\rho\Delta\epsilon$ and $\gamma_p=(\Delta\epsilon/p\pi)\sin(p\pi\rho)$, where $\Delta\epsilon=\epsilon_h-\epsilon_a$.

At normal incidence, the Bloch wavevector along the $x$ direction is zero, and the electric field can be expanded exactly as
\begin{equation}
E_y(x,z)=\sum_{m=-\infty}^{\infty}E_m(z)e^{imKx},
\label{eq:bloch}
\end{equation}
where $E_m(z)$ is the amplitude of the $m$th spatial harmonic. Substituting Eqs.~(\ref{eq:eps}) and (\ref{eq:bloch}) into Eq.~(\ref{eq:wave}) yields
\begin{equation}
\frac{d^{2}E_m}{dz^{2}}+\left[k_0^2\gamma_0(z)-(mK)^2\right]E_m=-k_0^2\sum_{p\neq0}\gamma_p(z)E_{m-p}.
\label{eq:coupled}
\end{equation}
Equation~(\ref{eq:coupled}) is exact and does not assume weak dielectric modulation. Each term $\gamma_pE_{m-p}$ describes coupling from the $(m-p)$th spatial harmonic to the $m$th spatial harmonic through the $p$th Fourier component of the dielectric modulation, with an in-plane momentum transfer $pK$. Outside the grating layer, $\gamma_{p\neq0}=0$, and the diffraction orders are uncoupled. The $z$ components of the $m$th-order diffraction wavevectors in the substrate and air are $q_{s,m}=\sqrt{k_0^2\epsilon_s-(mK)^2}$ and $q_{a,m}=\sqrt{k_0^2\epsilon_a-(mK)^2}$, respectively. The $m$th diffraction order is propagating when the corresponding wavevector is real and evanescent when it is imaginary.

Separating the background from the coupling terms, Eq.~(\ref{eq:coupled}) can be written as
\begin{equation}
\mathcal{L}_m E_m(z)=F_m(z),
\label{eq:source}
\end{equation}
where the background operator is $\mathcal{L}_m=d^{2}/dz^{2}+k_0^2\gamma_0(z)-(mK)^2$ and the source term is $F_m(z)=-k_0^2\sum_{p\neq0}\gamma_p E_{m-p}(z)$. The source term $F_m(z)$ collects the modulation-induced coupling contributions that drive the $m$th spatial harmonic $E_m(z)$. At the grating interfaces, each spatial harmonic $E_m(z)$ matches the corresponding external diffraction order, which propagates when its channel is open and decays evanescently when it is closed.
For the laterally mirror-symmetric unit cell considered here, normal incidence excites the even symmetry sector, for which $E_{-m}(z)=E_m(z)$. Together with $\gamma_{-p}=\gamma_p$, the contributions from the paired spatial harmonics $+n$ and $-n$ can be combined, and for $m>0$ the source term becomes
\begin{equation}
\begin{split}
F_m(z)=-k_0^2\Bigl[&\gamma_m E_0(z)+\gamma_{2m}E_m(z)\\
&+\sum_{n\geq1,\,n\neq m}\left(\gamma_{|m-n|}+\gamma_{m+n}\right)E_n(z)\Bigr].
\end{split}
\label{eq:Fm}
\end{equation}
The coefficient $\gamma_{|m-n|}+\gamma_{m+n}$ accounts for the two coupling pathways into the $m$th spatial harmonic from the field components $E_{+n}(z)$ and $E_{-n}(z)$. The term $\gamma_{2m}E_m(z)$ originates from coupling from $E_{-m}(z)$ to $E_{+m}(z)$ through the $2m$th Fourier component of the dielectric modulation, where $E_{-m}(z)=E_m(z)$ has been used.

To relate the source to the amplitudes radiated from the grating, we introduce two homogeneous solutions of $\mathcal{L}_m$, $u_m$ and $v_m$, which satisfy the outgoing conditions into the substrate and air, respectively, and are normalized as $u_m(0)=1$ and $v_m(t)=1$ (Appendix~\ref{app:green}). Here, $k_m=\sqrt{k_0^2\gamma_0-(mK)^2}$ is the wavevector of the $m$th background harmonic inside the grating layer.

The incident wave excites a field inside the grating layer, and the modulation-induced source $F_m$ generated by this field produces outgoing waves in the corresponding $m$th diffraction channel. With the incident electric-field amplitude normalized to unity, we denote the resulting source-generated reflection and transmission amplitudes by $r_m$ and $t_m$, respectively,
\begin{align}
r_m&=\frac{1}{W_m}\int_{0}^{t}v_m(z)F_m(z)\,dz,
\label{eq:rm}\\
t_m&=\frac{1}{W_m}\int_{0}^{t}u_m(z)F_m(z)\,dz.
\label{eq:tm}
\end{align}
Here, $W_m=u_mv_m'-u_m'v_m$ is the Wronskian of the two homogeneous solutions. For $m\neq0$, no incident amplitude exists in the corresponding diffraction channel, and Eqs.~(\ref{eq:rm}) and (\ref{eq:tm}) therefore give the total outgoing amplitudes of that order. For a propagating order, the diffraction efficiencies are
\begin{equation}
R_m=\frac{q_{s,m}}{q_{s,0}}\left|r_m\right|^{2},
\qquad
T_m=\frac{q_{a,m}}{q_{s,0}}\left|t_m\right|^{2}.
\label{eq:RTm}
\end{equation}
The zeroth-order channel is different because the incident wave also excites a background field that remains finite even in the absence of dielectric modulation. The zeroth-order reflectance and transmittance are therefore
\begin{equation}
R_0=\left|r_0^{\mathrm{bg}}+r_0\right|^{2},
\qquad
T_0=\frac{q_{a,0}}{q_{s,0}}\left|t_0^{\mathrm{bg}}+t_0\right|^{2},
\label{eq:RT0}
\end{equation}
where $r_0^{\mathrm{bg}}$ and $t_0^{\mathrm{bg}}$ are the reflection and transmission amplitudes of a uniform slab of dielectric constant $\gamma_0$ and thickness $t$ between the substrate and air (Appendix~\ref{app:derivation}).

Equations~(\ref{eq:rm}) and (\ref{eq:tm}) do not by themselves determine the diffraction amplitudes, because the source $F_m$ is constructed from the spatial harmonics $E_m(z)$, which must first be obtained from the coupled equations. Their role here is instead to provide a channel-resolved interpretation of the self-consistent solution. Once the spatial harmonics are known, each source-generated amplitude can be expressed as a source integral whose individual coupling contributions can be examined separately. In the calculations below, the spatial harmonics are obtained from an RCWA solver implemented for this work, with the field expansion truncated at $|m|\leq32$. The total diffraction amplitudes reconstructed using Eqs.~(\ref{eq:rm}) and (\ref{eq:tm}), with the background contributions added for the zeroth order, agree with those obtained directly from RCWA (Appendix~\ref{app:numerics}).

\begin{figure*}[t]
\centering
\includegraphics[width=\textwidth]{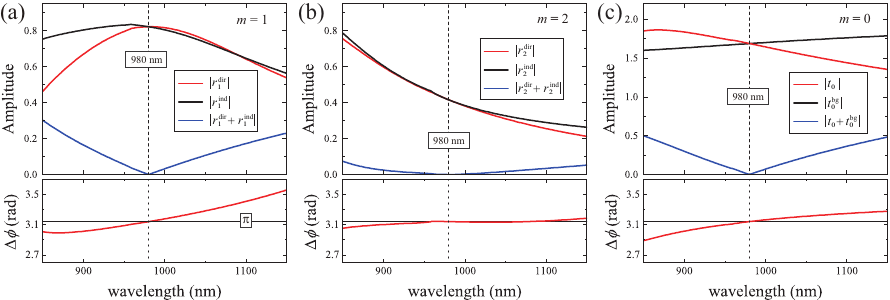}
\caption{Channel-resolved amplitude decomposition for the TE-optimized MHCG of Refs.~\cite{Gebski2015,Dems2017}. (a), (b) Magnitudes of the direct, indirect, and total reflection amplitudes for the $m=1$ and $m=2$ diffraction channels, respectively. The lower panels show the corresponding phase difference $\Delta\phi$ between the direct and indirect contributions. (c) Magnitudes of the modulation-induced, background, and total zeroth-order transmission amplitudes, with the phase difference between the modulation-induced and background contributions shown in the lower panel. The vertical dashed lines mark $980$~nm, where the two contributing amplitudes in each panel have equal magnitudes and $\Delta\phi=\pi$, resulting in a vanishing total amplitude. The horizontal lines in the lower panels indicate $\Delta\phi=\pi$. Lateral mirror symmetry at normal incidence gives identical results for the corresponding negative diffraction orders.}
\label{fig2}
\end{figure*}

\section{Direct and indirect contributions to the diffracted amplitudes}
\label{sec:decomp}

For $m>0$, Eq.~(\ref{eq:Fm}) separates the source driving the $m$th spatial harmonic into the direct pathway from $E_0$ and the remaining Fourier-mediated pathways. We define the direct contribution as
\begin{equation}
F_m^{\mathrm{dir}}(z)=-k_0^2\gamma_mE_0(z),
\label{eq:Fdir}
\end{equation}
which describes coupling from the zeroth spatial harmonic $E_0$ to the $m$th spatial harmonic through the $m$th Fourier component $\gamma_m$. Here, direct refers only to this coupling step in the source, and $E_0$, like the other spatial harmonics, is obtained self-consistently from the full set of coupled equations, so that the direct contribution is not a single-scattering approximation. The remaining terms define the indirect contribution,
\begin{equation}
\begin{split}
F_m^{\mathrm{ind}}(z)=-k_0^2\Bigl[\gamma_{2m}&E_m(z)\\
+\sum_{n\geq1,\,n\neq m}&\bigl(\gamma_{|m-n|}+\gamma_{m+n}\bigr)E_n(z)\Bigr],
\end{split}
\label{eq:Find}
\end{equation}
so that $F_m=F_m^{\mathrm{dir}}+F_m^{\mathrm{ind}}$. Because Eqs.~(\ref{eq:rm}) and (\ref{eq:tm}) are linear in $F_m$, the same decomposition carries directly over to the reflected amplitude,
\begin{equation}
r_m=r_m^{\mathrm{dir}}+r_m^{\mathrm{ind}},
\label{eq:rsplit}
\end{equation}
and likewise to the transmitted amplitude $t_m$. Since the direct and indirect contributions generate outgoing waves in the same $m$th diffraction channel, their complex amplitudes interfere.

A higher-order diffraction channel is extinguished when its direct and indirect outgoing amplitudes cancel. For a substrate-side channel, $r_m=0$ requires
\begin{equation}
\left|r_m^{\mathrm{dir}}\right|=\left|r_m^{\mathrm{ind}}\right|,
\qquad
\arg\left(r_m^{\mathrm{dir}}\right)-\arg\left(r_m^{\mathrm{ind}}\right)=\pi,
\label{eq:cancelcond}
\end{equation}
where the phase difference is defined modulo $2\pi$. Importantly, this cancellation can occur even when the corresponding spatial harmonic $E_m(z)$ remains finite inside the grating, because the direct and indirect radiative contributions cancel in the outgoing channel.
The zeroth-order channel requires a separate condition because it contains the incident-wave background. For $m=0$, the average term $\gamma_0E_0$ belongs to the background operator $\mathcal{L}_0$, leaving the modulation-induced source
\begin{equation}
F_0(z)=-2k_0^2\sum_{n\geq1}\gamma_nE_n(z).
\label{eq:F0}
\end{equation}
The total zeroth-order transmission vanishes when
\begin{equation}
t_0^{\mathrm{bg}}+t_0=0.
\label{eq:t0cancel}
\end{equation}
For a lossless structure, cancellation of the zeroth-order transmission together with the extinction of all open higher-order channels leaves the zeroth-order reflected wave as the only propagating output channel, yielding $R_0=1$.

We now apply the decomposition to the TE-optimized MHCG reported in Refs.~\cite{Gebski2015,Dems2017}. Figures~\ref{fig2}(a) and \ref{fig2}(b) show the magnitudes of the direct, indirect, and total reflection amplitudes for the $m=1$ and $m=2$ diffraction channels, respectively, together with their phase difference $\Delta\phi$. At 980~nm, the direct and indirect contributions remain finite and have equal magnitudes, $|r_1^{\mathrm{dir}}|=|r_1^{\mathrm{ind}}|=0.822$ and $|r_2^{\mathrm{dir}}|=|r_2^{\mathrm{ind}}|=0.416$, while $\Delta\phi=\pi$ for both channels. The corresponding total amplitudes vanish, demonstrating that the open $m=\pm1$ and $m=\pm2$ channels satisfy the direct--indirect cancellation condition of Eq.~(\ref{eq:cancelcond}). Figure~\ref{fig2}(c) shows the corresponding decomposition of the zeroth-order transmission. Here, the interfering contributions are the background transmission amplitude $t_0^{\mathrm{bg}}$ and the modulation-induced amplitude $t_0$. At 980~nm, their magnitudes are equal, $|t_0^{\mathrm{bg}}|=|t_0|=1.69$, and $\Delta\phi=\pi$, causing the total transmission amplitude $t_0^{\mathrm{bg}}+t_0$ to vanish. Taken together, the amplitude decompositions in Fig.~\ref{fig2} provide direct numerical verification of the channel-resolved cancellation mechanism underlying the scatter-free perfect reflection.

\begin{figure*}[t]
\centering
\includegraphics[width=\textwidth]{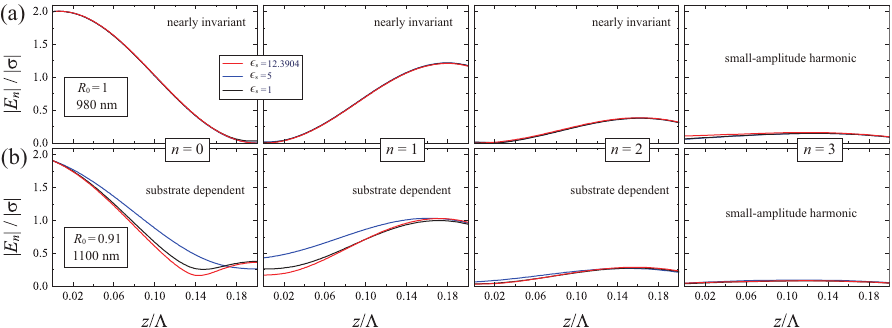}
\caption{Near invariance of the field harmonics under changes in the substrate dielectric constant. (a), (b) Magnitudes of the harmonics $n=0,1,2,3$ inside the grating layer, shown from left to right, at $980$~nm and $1100$~nm, respectively. The red, blue, and black curves correspond to $\epsilon_s=12.3904$, $5$, and $1$, respectively. The profiles for the changed substrates are normalized by $|\sigma|=|\tilde E_0(0)/E_0(0)|$, using $\epsilon_s=12.3904$ as the reference at each wavelength. At $980$~nm, the dominant harmonic profiles nearly overlap, whereas at $1100$~nm they exhibit a clear substrate dependence, most prominently for $n=0$ and $n=1$. For the reference substrate, the diffraction orders corresponding to $n=0,1,2$ are open in the substrate, whereas that corresponding to $n=3$ is closed. Harmonics with $n\geq4$ are smaller than the $n=3$ harmonic and are not shown. The reflectances $R_0=1$ and $R_0=0.91$ indicated in (a) and (b) refer to $\epsilon_s=12.3904$.}
\label{fig3}
\end{figure*}

\section{Stability of the cancellation conditions against the substrate}
\label{sec:substrate}

Figure~\ref{fig1}(c) shows that the scatter-free perfect reflection of the MHCG is remarkably stable against changes in the substrate dielectric constant. This stability has been attributed to the structure of the grating admittance matrix and to the nearly flat superposition of grating modes at the input interface~\cite{Dems2017,Pruszynska2023}. In the present description, the extinction of an open higher-order channel is a balance between its direct and indirect amplitudes, and both amplitudes are generated by the field harmonics inside the grating. In this section, we show that the field harmonics excited in the grating layer remain nearly unchanged as the substrate is varied, so that the cancellation conditions remain nearly satisfied and the zeroth-order reflectance stays nearly perfect.

Figure~\ref{fig3}(a) compares the field harmonics at $980$~nm for the reference substrate $\epsilon_s=12.3904$ and the changed substrates $\epsilon_s=5$ and $1$. To compare their spatial profiles, we normalize the changed fields by $\sigma=\tilde E_0(0)/E_0(0)$, where untilded and tilded quantities refer to the reference and changed substrates, respectively. The plotted profiles $|\tilde E_n(z)|/|\sigma|$ nearly overlap with $|E_n(z)|$ for $n=0,1,2$, whereas the $n=3$ harmonic remains small. The dominant field harmonics thus retain nearly the same magnitude profiles despite the large change in substrate dielectric constant. By contrast, Fig.~\ref{fig3}(b) shows the same harmonics at $1100$~nm, where the reference-substrate reflectance is $R_0=0.91$. After the same normalization, the profiles show a clear substrate dependence, most prominently for $n=0$ and $n=1$. The difference between the two wavelengths can be traced to the boundary conditions of the extinguished channels.

In Eq.~(\ref{eq:coupled}), the Fourier coefficients $\gamma_p$ and the wavevectors $k_m$ are independent of $\epsilon_s$. The outgoing conditions at the upper, air-side interface are also unchanged. Thus, the substrate dependence enters only through the lower-interface boundary conditions. For $m\neq0$, these conditions are
\begin{equation}
E_m(0)=r_m,\qquad E_m'(0)=-iq_{s,m}r_m,
\label{eq:lowerbc}
\end{equation}
where the prime denotes differentiation with respect to $z$. At the scatter-free point, $r_m=0$ for each open higher-order channel, so both $E_m(0)$ and $E_m'(0)$ vanish. These zero boundary values remain compatible with the boundary condition when $q_{s,m}$ changes. Channel extinction therefore removes the direct boundary forcing caused by the substrate change, even though the corresponding harmonic remains finite inside the grating. Writing the change after normalization as $\Delta_m(z)=\tilde E_m(z)-\sigma E_m(z)$ and subtracting the boundary conditions for the two substrates gives
\begin{equation}
\Delta_m'(0)+i\tilde q_{s,m}\Delta_m(0)=-i\sigma(\tilde q_{s,m}-q_{s,m})r_m,\qquad m\neq0.
\label{eq:substrateresidual}
\end{equation}
The right-hand side vanishes for the extinguished channels at $980$~nm, whereas it remains nonzero for $r_1$ and $r_2$ at $1100$~nm. A vanishing right-hand side does not require $\Delta_m=0$, because changes in other harmonics can still be transmitted through the coupled equations. In particular, closed orders with nonzero lower-interface amplitudes remain directly perturbed by the substrate. At the reference substrate, their contributions to the higher-order reflected amplitudes are relatively weak, as evaluated through the source integrals. This helps limit their effect on the cancellation between the dominant contributions.

The connection between the nearly preserved harmonics and the higher-order cancellation follows from Eq.~(\ref{eq:rm}). The projection kernel $v_m$ is unchanged because the grating and the air side are fixed. If the field harmonics scale by a common complex factor, the direct and indirect source integrals scale by that same factor. The substrate-dependent factor $1/W_m$ also multiplies both contributions equally. Thus, although the individual radiative amplitudes may change, their equal magnitudes and opposite phases are preserved, and their sum remains zero. The near invariance of the dominant harmonics therefore helps maintain the direct--indirect balance as the substrate changes.

The zeroth harmonic obeys a different condition because it contains the incident field, and its lower-interface field and derivative cannot vanish simultaneously (Appendix~\ref{app:zeroth}). The substrate change can alter the reflection phase and the overall field amplitude, while the normalized zeroth-harmonic profile remains nearly unchanged, as shown in Fig.~\ref{fig3}(a). At the fixed air-side interface, $E_0(t)=t_0^{\mathrm{bg}}+t_0$ remains close to zero, maintaining the cancellation between the background and modulation-induced transmitted fields. Together with the higher-order cancellations, this sustains the nearly perfect zeroth-order reflectance. This mechanism also applies as substrate diffraction orders close. Closing an order removes its radiative loss channel but leaves its spatial harmonic in the coupled equations. For an order extinguished at the reference substrate, the direct boundary forcing in Eq.~(\ref{eq:substrateresidual}) remains zero across the cutoff, consistent with the stability extending across the channel thresholds in Fig.~\ref{fig1}(c).

\begin{figure*}[t]
\centering
\includegraphics[width=14cm]{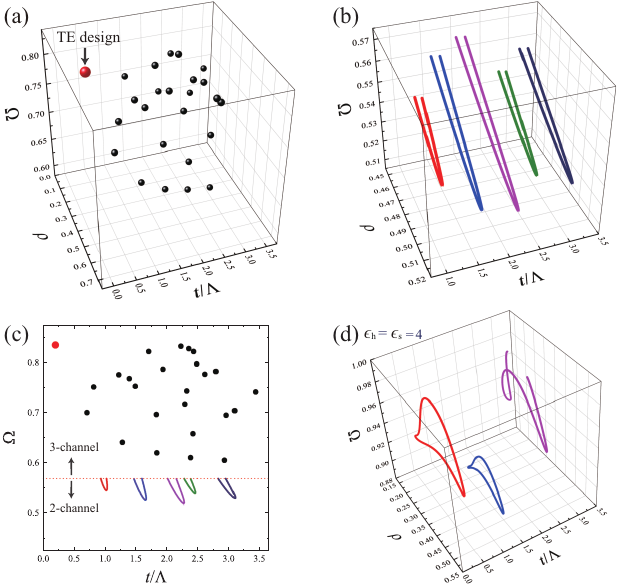}
\caption{Perfect-reflection solutions of the monolithic grating in the space of the design parameters $(\rho,t/\Lambda,\Omega)$. (a) Isolated solutions in the three-channel window for $\epsilon_h=\epsilon_s=12.3904$. The red sphere marks the TE design of Refs.~\cite{Gebski2015,Dems2017}. (b) Solution branches in the two-channel window for the same dielectric constants. (c) Solutions of (a) and (b) projected onto the $(t/\Lambda,\Omega)$ plane. The red dotted line marks $\Omega=2/\sqrt{\epsilon_s}$, the Rayleigh threshold at which the substrate $\pm2$ orders open, and separates the three-channel window above it from the two-channel window below it. (d) Solutions in the two-channel window for $\epsilon_h=\epsilon_s=4$. Two branches form closed loops within the window, whereas the third reaches the upper boundary $\Omega=1$ at both ends, where the substrate $\pm2$ and air-side $\pm1$ orders open simultaneously. In (b)--(d), different colors distinguish separate branches.}
\label{fig4}
\end{figure*}

\section{Channel counting and dimensionality of the perfect-reflection solutions}
\label{sec:counting}

The decomposition in Sec.~\ref{sec:decomp} shows that perfect reflection requires the extinction of all open higher-order channels and the cancellation of the zeroth-order transmission. At normal incidence, mirror symmetry gives $E_{-m}=E_m$, so that each pair of orders $\pm m$ contributes one cancellation condition, $r_m=0$. Let $N$ denote the total number of symmetry-compatible open channels, including the incident channel. Although the cancellation conditions involve complex amplitudes, their real and imaginary parts cannot all be counted as independent conditions. Losslessness and reciprocity constrain the amplitudes through a symmetric and unitary scattering matrix~\cite{Haus1984,Fan2003}. As a result, perfect reflection generically requires $N-1$ independent real conditions (Appendix~\ref{app:counting}). Here, the number $N-1$ is the generic codimension of the perfect-reflection solution set, that is, the number of dimensions by which these conditions reduce the space of design parameters. Let $P$ denote the number of independent design parameters available to satisfy these conditions. For the standard single-ridge MHCG, we take $\rho$, $t/\Lambda$, and the normalized frequency $\Omega=\Lambda/\lambda$ as the design parameters, giving $P=3$.

In the three-channel window, $2/\sqrt{\epsilon_s}<\Omega<3/\sqrt{\epsilon_s}$, the generic codimension of the perfect-reflection solution set is $N-1=3$, since the substrate orders $0$, $\pm1$, and $\pm2$ and the zeroth-order transmission into air give $N=4$. With $P=N-1$, no continuous design freedom remains after satisfying the cancellation conditions, and the perfect-reflection solutions are generically isolated points. Figure~\ref{fig4}(a) shows the $25$ isolated perfect-reflection solutions found for $\epsilon_h=\epsilon_s=12.3904$ using the numerical search described in Appendix~\ref{sm:tworidge}. The isolated nature of these solutions is consistent with the predicted dimensionality and with the separate reflectance maxima found in Ref.~\cite{Marciniak2016} by optimizing the period, ridge width, and grating height. The red sphere in Fig.~\ref{fig4}(a) corresponds to the TE design of Refs.~\cite{Gebski2015,Dems2017}.

\begin{figure*}[t]
\centering
\includegraphics[width=\textwidth]{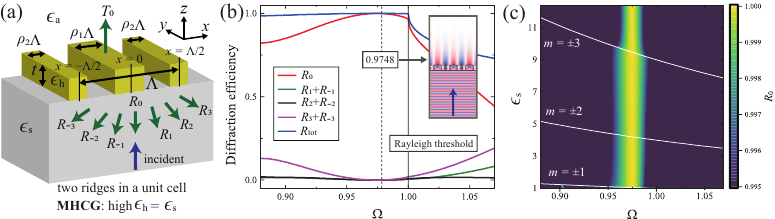}
\caption{Perfect reflection in the four-channel window restored by an additional design parameter. (a) Schematic of the MHCG with two ridges of widths $\rho_1\Lambda$ and $\rho_2\Lambda$ centered at $x=0$ and on the unit-cell boundary $x=\pm\Lambda/2$, and the propagating diffraction channels for normal incidence from the substrate. At the perfect-reflection frequency, the air side supports only the zeroth transmitted order, whereas the substrate supports the reflected orders $m=0,\pm1,\pm2,\pm3$. (b), (c) RCWA results for a selected solution, with $\rho_1=0.4537$, $\rho_2=0.0677$, $t/\Lambda=0.4328$, and $\epsilon_h=12.3904$. (b) Diffraction efficiencies of the reflected orders as a function of $\Omega$ for $\epsilon_s=\epsilon_h$. The dashed line marks $\Omega=0.9748$, where the higher-order reflection vanishes and $R_0$ reaches unity. $R_{\mathrm{tot}}$ is the total reflected efficiency, and the vertical solid line marks the Rayleigh threshold $\Omega=1$, where the air-side $\pm1$ orders open. The inset shows the real part of $E_y$ at $\Omega=0.9748$ obtained by the FEM simulation for the incident wave entering from the substrate. (c) Zeroth-order efficiency $R_0$ as a function of $\Omega$ and $\epsilon_s$ for the same structure. The white lines mark the Rayleigh thresholds $\epsilon_s=(m/\Omega)^2$ of the substrate orders $m=\pm1$, $\pm2$, and $\pm3$.}
\label{fig5}
\end{figure*}

In the two-channel window, $1/\sqrt{\epsilon_s}<\Omega<2/\sqrt{\epsilon_s}$, the generic codimension of the perfect-reflection solution set is $N-1=2$, since the substrate orders $0$ and $\pm1$ and the zeroth-order transmission into air give $N=3$. Because $P$ exceeds $N-1$ by one, one continuous degree of design freedom remains after satisfying the cancellation conditions, and the perfect-reflection solutions generically form curves. Figure~\ref{fig4}(b) shows the five separate branches found for the same dielectric constants, all within $0.93<t/\Lambda<3.10$. Further branches may appear at larger thicknesses. To our knowledge, the two-channel perfect-reflection solutions of MHCGs have not been reported as continuous curves.

All five branches in Fig.~\ref{fig4}(b) are arcs cut at both ends by the Rayleigh threshold of the substrate $\pm2$ orders~\cite{Rayleigh1907,Hessel1965}, $\Omega=2/\sqrt{\epsilon_s}=0.5682$, as seen in Fig.~\ref{fig4}(c). Above this threshold, the $\pm2$ orders propagate in the substrate, and perfect reflection requires the additional cancellation condition $r_2=0$, which the branches do not satisfy. The solution branches can also form closed loops within the two-channel window. For an MHCG with $\epsilon_h=\epsilon_s=4$, Fig.~\ref{fig4}(d) shows two closed loops in red and blue, while the magenta branch reaches the upper boundary $\Omega=2/\sqrt{\epsilon_s}=1$ at both ends.

Returning to the MHCG with $\epsilon_h=\epsilon_s=12.3904$, we consider the four-channel window, $3/\sqrt{\epsilon_s}<\Omega<1$, where the substrate $\pm3$ orders also propagate. This window lies beyond the two- and three-channel regimes of the TM and TE designs reported in previous studies, respectively~\cite{Gebski2015,Dems2017}. The air side still supports only the zeroth transmitted order, giving $N=5$ and a generic codimension of $N-1=4$. With $P=3<N-1$, there are more independent conditions than design parameters, so no perfect-reflection solution is generically expected. Our numerical search found none in this window (Appendix~\ref{sm:tworidge}). Nevertheless, this counting does not exclude accidental solutions, for which all four conditions happen to be satisfied simultaneously despite the limited number of design parameters.

Increasing $P$ to match the generic codimension $N-1=4$ makes isolated perfect-reflection solutions generically possible. We therefore introduce a second ridge of width $\rho_2\Lambda$ centered on the unit-cell boundary $x=\pm\Lambda/2$, while the ridge at $x=0$ has width $\rho_1\Lambda$, as shown in Fig.~\ref{fig5}(a). The mirror symmetry about $x=0$ is preserved, so the orders $\pm m$ remain paired and the channel count is unchanged. The four independent design parameters are now $\rho_1$, $\rho_2$, $t/\Lambda$, and $\Omega$, giving $P=N-1=4$. Our numerical search found $15$ isolated perfect-reflection solutions, listed in Table~\ref{tab:S1} in Appendix~\ref{sm:tworidge}. The use of an additional structural element to control more diffraction channels is conceptually related to metagratings, in which the degrees of freedom within each period are used to control the amplitudes of propagating diffraction orders~\cite{Radi2017,Popov2018,Wong2018}.

Figure~\ref{fig5}(b) shows the diffraction efficiencies of a selected solution, with $\rho_1=0.4537$, $\rho_2=0.0677$, and $t/\Lambda=0.4328$, calculated by RCWA. At $\Omega=0.9748$, the reflected efficiencies of the $\pm1$, $\pm2$, and $\pm3$ orders and the zeroth-order transmission vanish simultaneously, and the zeroth-order reflectance reaches unity. The field distribution in the inset, obtained by FEM, shows that away from the grating only the incident and zeroth-order reflected waves remain in the substrate. Figure~\ref{fig5}(c) shows that the zeroth-order reflectance remains nearly perfect as $\epsilon_s$ is reduced from $12.3904$ to $1$, even as the substrate $\pm3$, $\pm2$, and $\pm1$ orders close successively at the Rayleigh thresholds marked by the white lines. Thus, the additional ridge enables perfect reflection with more open diffraction channels while preserving the remarkable substrate stability, consistent with the mechanism explained in Sec.~\ref{sec:substrate}.

\begin{figure*}[t]
\centering
\includegraphics[width=14cm]{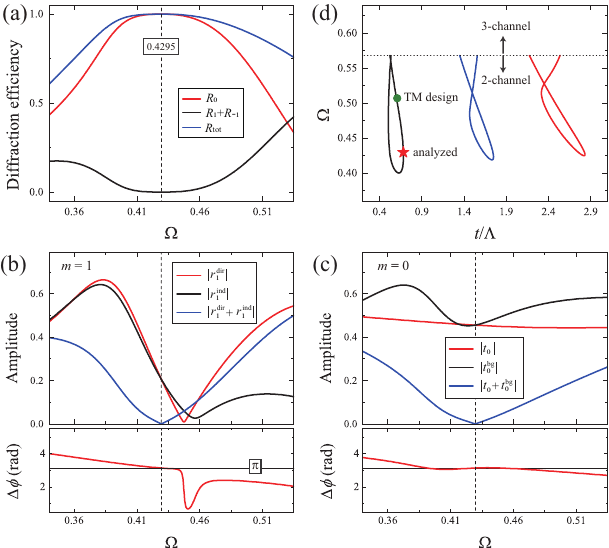}
\caption{TM perfect reflection of the monolithic grating with $\epsilon_h=\epsilon_s=12.3904$. (a) Diffraction efficiencies as a function of $\Omega$ for $\rho=0.5703$ and $t/\Lambda=0.6813$, calculated by RCWA. $R_{\mathrm{tot}}$ is the total reflected efficiency, and the dashed line marks the perfect-reflection frequency $\Omega=0.4295$. (b) Magnitudes of the direct, indirect, and total $m=1$ reflected amplitudes, with the phase difference $\Delta\phi$ between the direct and indirect contributions shown in the lower panel. (c) Magnitudes of the modulation-induced, background, and total zeroth-order transmission amplitudes, with the phase difference between the modulation-induced and background contributions shown in the lower panel. The vertical dashed lines in (b) and (c) mark $\Omega=0.4295$, where the two contributing amplitudes in each panel have equal magnitudes and $\Delta\phi=\pi$, resulting in a vanishing total amplitude. The horizontal lines in the lower panels indicate $\Delta\phi=\pi$. (d) Perfect-reflection solutions in the two-channel window projected onto the $(t/\Lambda,\Omega)$ plane. Different colors distinguish separate branches. The dotted line marks $\Omega=2/\sqrt{\epsilon_s}$, the Rayleigh threshold at which the substrate $\pm2$ orders open, and separates the three-channel window above it from the two-channel window below it. The red star marks the solution analyzed in (a)--(c), and the green circle marks the TM design of Ref.~\cite{Dems2017}.}
\label{fig6}
\end{figure*}

\section{Extension to TM polarization}
\label{sec:tm}

High-reflectivity MHCGs have also been designed for TM polarization, and the absence of scattering in them was analyzed in terms of the grating impedance matrix~\cite{Dems2017}. In this section, we show that the cancellation mechanisms and channel counting established for TE also apply to TM polarization. For this polarization, the magnetic field $H_y=\sum_mH_m(z)e^{imKx}$ is governed by the Fourier components $\eta_p$ of $1/\epsilon$. The coupled equations are constructed using the Fourier factorization rules~\cite{Li1996,Lalanne1996,Granet1996}. As for TE, separating the background terms gives $\mathcal{L}_mH_m=F_m$, with the TM background layer and source given in Appendix~\ref{sm:tm}. An additional difference from TE arises at the interfaces, where $\epsilon^{-1}\partial_zH_y$ rather than $\partial_zH_y$ is continuous. The $m$th Fourier component of $\epsilon^{-1}\partial_zH_y$ couples the derivatives $H_n'$ of all harmonics through $\eta_{m-n}$. The background boundary condition retains only the diagonal term $\eta_0H_m'$, while the remaining couplings define the boundary mismatch $S_m=\eta_0^{-1}\sum_{n\neq m}\eta_{m-n}H_n'$. The reflected amplitude therefore acquires an interface term in addition to the volume integral,
\begin{multline}
r_m=\frac{1}{W_m}\int_0^t v_m(z)F_m(z)\,dz\\
+\frac{1}{W_m}\bigl[S_m(t)-v_m(0)S_m(0)\bigr],
\label{eq:tmrm}
\end{multline}
where $v_m$ and $W_m$ are now defined for the TM background layer. For $m\neq0$, the $n=0$ terms of $F_m$ and $S_m$ form the direct contribution and the remaining terms the indirect contribution, so that the decomposition of Eq.~(\ref{eq:rsplit}) and the cancellation condition of Eq.~(\ref{eq:cancelcond}) apply unchanged.

Figure~\ref{fig6}(a) shows the diffraction efficiencies of a TM perfect-reflection solution in the two-channel window with $\rho=0.5703$ and $t/\Lambda=0.6813$. The zeroth-order efficiency reaches unity at $\Omega=0.4295$, where the $\pm1$ efficiencies and the zeroth-order transmission vanish. Figure~\ref{fig6}(b) shows that, at $\Omega=0.4295$, the direct and indirect contributions to the $m=1$ reflected amplitude have equal magnitudes, $|r_1^{\mathrm{dir}}|=|r_1^{\mathrm{ind}}|=0.206$, and a phase difference $\Delta\phi=\pi$. Figure~\ref{fig6}(c) shows that the zeroth-order transmission vanishes through the same cancellation between the background and modulation-induced amplitudes as for TE, with equal magnitudes $|t_0^{\mathrm{bg}}|=|t_0|=0.455$ and a phase difference $\Delta\phi=\pi$.

The counting of Sec.~\ref{sec:counting} relies only on losslessness, reciprocity, and the symmetry of the unit cell, and therefore applies equally to TM polarization. In the two-channel window, the TM solutions form curves, and three separate branches were found within $0.50<t/\Lambda<2.84$, as shown in Fig.~\ref{fig6}(d). Further branches may appear at larger thicknesses. Both ends of each branch reach the boundary $\Omega=2/\sqrt{\epsilon_s}$. The solution analyzed in Figs.~\ref{fig6}(a)--\ref{fig6}(c) is marked by the red star on the black branch in Fig.~\ref{fig6}(d). In the three-channel window, the TM solutions appear as isolated points, and the ten solutions identified by our numerical search are listed in Table~\ref{tab:TM} in Appendix~\ref{sm:tworidge}. The TE and TM designs of Ref.~\cite{Dems2017} lie in the three- and two-channel windows, respectively. The TE design is an isolated point marked by the red sphere in Fig.~\ref{fig4}(a), whereas the TM design is marked by the green circle on the same black branch in Fig.~\ref{fig6}(d).

\section{Conclusion}

In conclusion, we have elucidated the mechanism of scatter-free perfect reflection in multichannel MHCGs using a Fourier-harmonic description. Open higher-order diffraction channels are extinguished by destructive interference between the radiation fields generated by the direct and indirect source contributions, while the zeroth-order transmission vanishes through cancellation between the background and modulation-induced amplitudes. Together, these cancellations yield scatter-free perfect reflection in lossless MHCGs. The near invariance of the field harmonics within the grating layer under changes in the substrate keeps these cancellation conditions nearly satisfied and accounts for the remarkable substrate stability of the perfect-reflection state. The number of symmetry-compatible open channels further determines the generic codimension of the perfect-reflection solution set. For the standard single-ridge MHCG with three independent design parameters $(\rho, t/\Lambda, \Omega)$, perfect-reflection solutions form continuous curves when the substrate supports only the zeroth and $\pm1$ orders, but become isolated points when the $\pm2$ orders also open. The opening of the higher-order $\pm3$ substrate channels increases the number of independent conditions to four, exceeding the three design parameters, so that no solution is generically expected. Introducing a second ridge increases the number of independent design parameters to four $(\rho_1, \rho_2, t/\Lambda, \Omega)$, and isolated perfect-reflection solutions are recovered numerically. Thus, the balance between the number of independent conditions and the number of design parameters determines the generic dimensionality of the perfect-reflection solution set. The same cancellation mechanisms and channel counting apply to both TE and TM polarizations. Our results connect the physical origin of scatter-free perfect reflection with the constraints imposed by multiple open diffraction channels, providing a unified understanding of its mechanism and solution structure.

\bigskip
\noindent\textbf{Acknowledgements}
This research was supported by a grant from the National Research Foundation of Korea, funded by the Ministry of Science and ICT (No. 2022R1A2C1011091).

\bigskip
\noindent\textbf{Conflict of Interest}\\
The authors declare no conflicts of interest.

\bigskip
\noindent\textbf{Data Availability Statement}\\
Data underlying the results in this paper may be obtained from the authors upon reasonable request.

\appendix
\renewcommand{\theequation}{\thesection\arabic{equation}}
\renewcommand{\thesubsection}{\thesection.\arabic{subsection}}
\makeatletter\renewcommand{\p@subsection}{}\makeatother

\section{Source form of the coupled equations and the Green-function projection}
\setcounter{equation}{0}
\label{app:derivation}

\subsection{Pairing of the orders $\pm m$}

For a real dielectric profile that is mirror symmetric about $x=0$, normal incidence gives $E_{-m}(z)=E_m(z)$ and real Fourier coefficients satisfying $\gamma_{-p}=\gamma_p$. Setting $n=m-p$ in the coupling sum of Eq.~(\ref{eq:coupled}), so that $p\neq0$ becomes $n\neq m$,
\begin{equation}
-k_0^2\sum_{p\neq0}\gamma_pE_{m-p}=-k_0^2\sum_{n\neq m}\gamma_{m-n}E_n .
\label{eq:app-shift}
\end{equation}
The term $n=0$ is $\gamma_mE_0$. For $m>0$, folding the negative indices with $n\to-n$ and using $E_{-n}=E_n$ turns $\sum_{n\leq-1}\gamma_{m-n}E_n$ into $\sum_{n\geq1}\gamma_{m+n}E_n$, whose $n=m$ member is $\gamma_{2m}E_m$. Collecting the three groups gives Eq.~(\ref{eq:Fm}). The pairing is exact and holds for arbitrary modulation strength. For example, for $m=2$ the source reads
\begin{equation}
F_2(z)=-k_0^2\bigl[\gamma_2E_0+(\gamma_1+\gamma_3)E_1+\gamma_4E_2+(\gamma_1+\gamma_5)E_3+\cdots\bigr],
\label{eq:app-F2}
\end{equation}
so that $r_2^{\mathrm{dir}}$ is driven by $\gamma_2E_0$ alone while $r_2^{\mathrm{ind}}$ collects every remaining term.

\subsection{Outgoing Green function}
\label{app:green}

Inside the layer $\gamma_0$ is constant, so $\mathcal{L}_m$ has constant coefficients, and its two outgoing solutions are
\begin{align}
u_m(z)&=\cos k_mz-i\frac{q_{s,m}}{k_m}\sin k_mz,
\label{eq:um}\\
v_m(z)&=\cos k_m(t-z)-i\frac{q_{a,m}}{k_m}\sin k_m(t-z).
\label{eq:vm}
\end{align}
Their Wronskian $W_m$ is independent of $z$, and the Green function of $\mathcal{L}_m$ with outgoing conditions at both faces is
\begin{equation}
G_m(z,z')=\frac{u_m(z_<)\,v_m(z_>)}{W_m} ,
\label{eq:app-green}
\end{equation}
with $z_<$ and $z_>$ the smaller and the larger of $z$ and $z'$. For $m\neq0$ there is no incident amplitude in that channel, so
\begin{equation}
E_m(z)=\int_{0}^{t}G_m(z,z')F_m(z')\,dz'
\label{eq:app-Em}
\end{equation}
is the complete solution. Evaluating it at the two faces and using $u_m(0)=1$ and $v_m(t)=1$ gives $r_m=E_m(0)$ and $t_m=E_m(t)$, which are Eqs.~(\ref{eq:rm}) and (\ref{eq:tm}). For an open substrate channel, $q_{s,m}$ is real, so $u_m$ and $W_m$ are complex and $r_m$ carries a phase, whereas for a closed substrate channel $q_{s,m}$ is purely imaginary and both become real. The air-side orders $m\neq0$ are closed in the windows considered, so $v_m$ is real in both cases.

\subsection{Zeroth-order background}

The zeroth-order background amplitudes are those of a uniform slab of dielectric constant $\gamma_0$ and thickness $t$. With $n_g=\sqrt{\gamma_0}$, $n_s=\sqrt{\epsilon_s}$, $n_a=\sqrt{\epsilon_a}$ and $\beta=k_0n_gt$,
\begin{align}
r_0^{\mathrm{bg}}&=\frac{r_{sg}+r_{ga}e^{2i\beta}}{1+r_{sg}r_{ga}e^{2i\beta}},
\label{eq:app-rbg}\\
t_0^{\mathrm{bg}}&=\frac{t_{sg}t_{ga}e^{i\beta}}{1+r_{sg}r_{ga}e^{2i\beta}},
\label{eq:app-tbg}
\end{align}
where $r_{sg}=(n_s-n_g)/(n_s+n_g)$, $r_{ga}=(n_g-n_a)/(n_g+n_a)$, $t_{sg}=2n_s/(n_s+n_g)$ and $t_{ga}=2n_g/(n_g+n_a)$. These amplitudes depend on the modulation only through $\gamma_0$, are independent of the Fourier components $\gamma_{p\neq0}$, and survive the limit $\Delta\epsilon\rightarrow0$, in which $F_0$ vanishes.

\subsection{Power balance}

For a lossless grating illuminated from the substrate,
\begin{equation}
R_0+\sum_{\substack{m\neq0\\ m\in\mathcal{O}_s}}R_m+\sum_{m\in\mathcal{O}_a}T_m=1 ,
\label{eq:app-power}
\end{equation}
where $\mathcal{O}_s$ and $\mathcal{O}_a$ are the sets of open diffraction orders in the substrate and in the air, so that the extinction of every open higher-order channel, Eq.~(\ref{eq:cancelcond}), together with $t_0^{\mathrm{bg}}+t_0=0$ gives $R_0=1$.

\subsection{Zeroth-order boundary condition and exact scaling}
\label{app:zeroth}

The zeroth harmonic contains the incident field. In terms of the total zeroth-order reflection amplitude $r_0^{\mathrm{tot}}=r_0^{\mathrm{bg}}+r_0$ of Eq.~(\ref{eq:RT0}), its lower-interface conditions are
\begin{equation}
E_0(0)=1+r_0^{\mathrm{tot}},\qquad E_0'(0)=iq_{s,0}\bigl(1-r_0^{\mathrm{tot}}\bigr).
\label{eq:lowerbc0}
\end{equation}
At the reference perfect-reflection point, $r_0^{\mathrm{tot}}=e^{i\varphi}$, and Eq.~(\ref{eq:lowerbc0}) gives the real ratio $E_0'(0)/E_0(0)=q_{s,0}\tan(\varphi/2)$. If $r_m=0$ for every $m\neq0$, including the evanescent orders, the scaled field $\sigma E_m$ satisfies all higher-order conditions of Eq.~(\ref{eq:lowerbc}) at the new substrate and satisfies Eq.~(\ref{eq:lowerbc0}) with $\tilde r_0^{\mathrm{tot}}=e^{i\tilde\varphi}$ if
\begin{equation}
\tilde q_{s,0}\tan\frac{\tilde\varphi}{2}=q_{s,0}\tan\frac{\varphi}{2},
\label{eq:exactscaling}
\end{equation}
which determines a real $\tilde\varphi$ for real positive $q_{s,0}$ and $\tilde q_{s,0}$. The corresponding scaling factor is $\sigma=(1+e^{i\tilde\varphi})/(1+e^{i\varphi})$, so that the scaled field also satisfies the boundary value $\tilde E_0(0)=1+e^{i\tilde\varphi}$. Because the coupled equations and the air-side conditions are unchanged, the scaled field is then an exact solution, and perfect reflection is exactly substrate independent. In the actual grating, the nonzero lower-interface amplitudes of the closed orders prevent this exact scaling.

\subsection{Numerical implementation and verification}
\label{app:numerics}

The RCWA solver uses the symmetry reduction $E_{-m}=E_m$ at normal incidence, and the Fourier coefficients required by the RCWA matrix are evaluated analytically from the exact binary dielectric profile. The solver reproduces the analytic reflectance of a uniform slab to within $6\times10^{-16}$, conserves power to within $2\times10^{-15}$, and agrees with an independent implementation to nine significant digits. Re-solving representative perfect-reflection solutions with $|m|\leq64$ and $128$ shifts them by at most $1.2\times10^{-5}$ in any design parameter. The source-integral representation of Eqs.~(\ref{eq:rm}) and (\ref{eq:tm}) is an identity that follows from the Green-function derivation of Appendix~\ref{app:green}. When the truncated self-consistent harmonics are inserted into these equations, the reconstructed higher-order diffraction amplitudes agree with those obtained directly from the RCWA to within $10^{-9}$ over the wavelength range considered, and the zeroth-order amplitudes agree to within $10^{-8}$ once the background amplitudes of Eq.~(\ref{eq:RT0}) are included.

\section{Scattering-matrix counting of the perfect-reflection conditions}
\setcounter{equation}{0}
\label{app:counting}

\subsection{Codimension of the perfect-reflection solution set}
\label{app:codim}

At normal incidence with a unit cell symmetric about $x=0$, the field satisfies $E_{-m}=E_m$ (Appendix~\ref{app:derivation}). The antisymmetric combinations of the orders $\pm m$ are therefore not excited, and the scattering problem closes on the symmetric channels. We take as channels the incident zeroth order in the substrate, the symmetric combinations $(|{+m}\rangle+|{-m}\rangle)/\sqrt{2}$ of the open substrate orders, and the open orders on the air side, and we normalize each amplitude so that its squared magnitude is a power flux, multiplying the TE field amplitude by $\sqrt{q}$ and the TM field amplitude by $\sqrt{q/\epsilon}$, where $q$ is the $z$-directed wavevector and $\epsilon$ the dielectric constant of the medium. With $N$ such open channels, the outgoing amplitudes are related to the incoming ones by an $N\times N$ scattering matrix $S$. For a lossless grating, $S^{\dagger}S=I$. For a reciprocal grating with a real dielectric constant, $S^{T}=S$ in this normalization~\cite{Haus1984,Fan2003}.

Perfect reflection requires the incident channel, labeled $1$, to be an eigenchannel,
\begin{equation}
S\mathbf{e}_1=e^{i\varphi}\mathbf{e}_1 .
\label{eq:app-eigen}
\end{equation}
Because $S$ is symmetric and unitary, it can be written in the Cayley form~\cite{Newton1982}
\begin{equation}
S=(I+iK)(I-iK)^{-1},
\label{eq:app-cayley}
\end{equation}
where $K$ is real and symmetric. This form requires that $-1$ is not an eigenvalue of $S$, which can always be arranged by multiplying $S$ by a constant phase. The matrices $S$ and $K$ share their eigenvectors, so that Eq.~(\ref{eq:app-eigen}) is equivalent to $K\mathbf{e}_1=\kappa\mathbf{e}_1$ with real $\kappa$, that is,
\begin{equation}
K_{21}=K_{31}=\cdots=K_{N1}=0 .
\label{eq:app-kcond}
\end{equation}
Each of these $N-1$ conditions is a single real equation in the design parameters. When their Jacobian has full rank, the perfect-reflection solution set therefore has codimension $N-1$. In the two-, three-, and four-channel windows of Sec.~\ref{sec:counting}, $N=3$, $4$, and $5$, respectively.

Equation~(\ref{eq:app-eigen}) is also equivalent to $S_{j1}=0$ for all $j\neq1$, which are the extinction conditions $r_m=0$ of the open higher orders and the cancellation $t_0^{\mathrm{bg}}+t_0=0$ of the zeroth-order transmission. These complex conditions are therefore not independent. Taken together, they impose $N-1$ real constraints rather than $2(N-1)$. An individual condition $S_{j1}=0$ is not equivalent to the corresponding $K_{j1}=0$, and the equivalence holds only when all of them are satisfied. The channel normalization becomes singular where an order reaches grazing propagation, so that the counting applies within each window, away from its boundaries.

\subsection{Time reversal and the common phase of the harmonics}
\label{app:timerev}

For a real dielectric constant $\epsilon(x,z)$, the complex conjugate $E^{*}$ of a solution is also a solution, with the roles of the incoming and outgoing waves exchanged. In the perfect-reflection state of Eq.~(\ref{eq:app-eigen}), the incoming wave has unit amplitude in channel $1$, and the outgoing wave has amplitude $e^{i\varphi}$ in the same channel. The conjugate solution therefore has incoming amplitude $e^{-i\varphi}$ and outgoing amplitude $1$ in channel $1$. Evanescent components remain evanescent under conjugation. If the scattering problem has a unique solution, that is, if no bound state exists at the same frequency~\cite{Hsu2016}, the conjugate solution equals $e^{-i\varphi}$ times the original one, $E^{*}=e^{-i\varphi}E$. The field $e^{-i\varphi/2}E(x,z)$ is then real. Its Fourier harmonics satisfy $e^{-i\varphi/2}E_{-m}=(e^{-i\varphi/2}E_m)^{*}$, and combined with $E_{-m}=E_m$ this gives
\begin{equation}
e^{-i\varphi/2}E_m(z)\in\mathbb{R}
\label{eq:app-real}
\end{equation}
for every $m$. All harmonics therefore share a common global phase, up to a sign. At the three-channel solution corresponding to the TE design (Sec.~\ref{sec:counting}), the calculated harmonics are real to within $10^{-13}$ after removal of this phase. At $z=0$, Eq.~(\ref{eq:app-real}) reproduces $E_0(0)=1+e^{i\varphi}=2\cos(\varphi/2)\,e^{i\varphi/2}$ of Eq.~(\ref{eq:lowerbc0}). Because the Fourier coefficients are real, the direct and indirect sources share the common phase of the field harmonics, up to a sign. For each higher-order reflected channel, the projection kernel $v_m$ is real, as described in Appendix~\ref{app:green}, and the factor $1/W_m$ is common to both contributions. Their relative phase is therefore $0$ or $\pi$.

\section{TM counterpart of the direct and indirect decomposition}
\setcounter{equation}{0}
\label{sm:tm}

This appendix gives the TM counterpart of the Fourier-harmonic description of Sec.~\ref{sec:fourier} and of the direct and indirect decomposition of Sec.~\ref{sec:decomp}. The grating layer occupies $0<z<t$, the substrate of dielectric constant $\epsilon_s$ lies below it, the cover is air with $\epsilon_a=1$, and light is incident from the substrate at normal incidence.

\subsection{Wave equation and Fourier expansions}

For TM polarization the magnetic field has the single component $H_y$. Inside the grating layer, which is uniform along $z$, the wave equation reads
\begin{equation}
\frac{1}{\epsilon(x)}\frac{\partial^{2}H_y}{\partial z^{2}}
+\frac{\partial}{\partial x}\left[\frac{1}{\epsilon(x)}\frac{\partial H_y}{\partial x}\right]
+k_0^{2}H_y=0 .
\label{eq:Twave}
\end{equation}
The inverse dielectric constant is expanded in Fourier components,
\begin{gather}
\frac{1}{\epsilon(x)}=\sum_{p}\eta_p\,e^{ipKx},
\qquad
\eta_0=\frac{\rho}{\epsilon_h}+\frac{1-\rho}{\epsilon_a},
\nonumber\\
\eta_p=\left(\frac{1}{\epsilon_h}-\frac{1}{\epsilon_a}\right)\frac{\sin(p\pi\rho)}{p\pi}\quad(p\neq0),
\label{eq:Teta}
\end{gather}
where $K=2\pi/\Lambda$, and the field is expanded as $H_y(x,z)=\sum_mH_m(z)e^{imKx}$. Substitution into Eq.~(\ref{eq:Twave}) gives
\begin{equation}
\sum_{n}\eta_{m-n}\frac{d^{2}H_n}{dz^{2}}
-K^{2}\sum_{n}mn\left[\epsilon^{-1}\right]_{mn}H_n
+k_0^{2}H_m=0 .
\label{eq:Tcoupled}
\end{equation}
Equation~(\ref{eq:Tcoupled}) is the TM counterpart of Eq.~(\ref{eq:coupled}). It differs from the TE case in three respects. The coupling coefficients are those of $1/\epsilon$ rather than of $\epsilon$. The coupled equations contain both $H_n''$ and $H_n$ contributions, and these are represented by different matrices. The term carrying $H_n''$ is the product of $1/\epsilon$ with a quantity that is continuous across the vertical ridge--groove interfaces, and its coefficients are the Fourier components $\eta_{m-n}$. The term carrying $H_n$ originates from $\partial_x[\epsilon^{-1}\partial_xH_y]$, in which $1/\epsilon$ and $\partial_xH_y$ are discontinuous at the same points. According to Li's factorization rule~\cite{Li1996}, it must be represented by $[\epsilon^{-1}]$, the inverse of the Toeplitz matrix of $\epsilon$, rather than by the Toeplitz matrix of $1/\epsilon$. The two representations coincide only in the untruncated limit.

\subsection{Background operator and source}

Separating the term $n=m$, which carries $\eta_0$, and dividing by $\eta_0$ puts Eq.~(\ref{eq:Tcoupled}) in the form
\begin{gather}
\mathcal{L}_m H_m(z)=F_m(z),
\qquad
\mathcal{L}_m=\frac{d^{2}}{dz^{2}}+k_m^{2},
\nonumber\\
k_m^{2}=\frac{k_0^{2}-(mK)^{2}\left[\epsilon^{-1}\right]_{mm}}{\eta_0},
\label{eq:Tsource}
\end{gather}
with the source
\begin{multline}
F_m(z)=\frac{1}{\eta_0}\Biggl\{
K^{2}\sum_{n\neq m}mn\left[\epsilon^{-1}\right]_{mn}H_n(z)\\
-\sum_{n\neq m}\eta_{m-n}\frac{d^{2}H_n}{dz^{2}}\Biggr\}.
\label{eq:TF}
\end{multline}
For the zeroth order the factor $mnK^2$ vanishes, and the background of the zeroth-order channel is a uniform layer of dielectric constant
\begin{equation}
\epsilon_{\mathrm{eff}}=\frac{1}{\eta_0},
\label{eq:Teeff}
\end{equation}
the harmonic mean of the profile, whereas for TE it is the arithmetic mean $\gamma_0$. For $m\neq0$, the diagonal element $[\epsilon^{-1}]_{mm}$ enters the background wavevector.
\subsection{Outgoing solutions and interface terms}

The quantities continuous at an interface are $H_y$ and $\epsilon^{-1}\partial_zH_y$. The outgoing solutions of the background operator therefore carry the ratios of the dielectric constants,
\begin{align}
u_m(z)&=\cos k_mz-i\,\frac{\epsilon_{\mathrm{eff}}}{\epsilon_s}\frac{q_{s,m}}{k_m}\sin k_mz,
\label{eq:Tum}\\
v_m(z)&=\cos k_m(t-z)-i\,\frac{\epsilon_{\mathrm{eff}}}{\epsilon_a}\frac{q_{a,m}}{k_m}\sin k_m(t-z),
\label{eq:Tvm}
\end{align}
normalized as $u_m(0)=1$ and $v_m(t)=1$, with $q_{s,m}$ and $q_{a,m}$ as defined in Sec.~\ref{sec:fourier} and the Wronskian $W_m=u_mv_m'-u_m'v_m$.

Inside the layer, $\epsilon^{-1}\partial_zH_y$ has the $m$th component $\sum_n\eta_{m-n}H_n'$, which mixes the harmonics. Keeping only the $\eta_0$ part in the background boundary condition leaves the remainder
\begin{equation}
S_m(z)\equiv\frac{1}{\eta_0}\sum_{n\neq m}\eta_{m-n}\,\frac{dH_n}{dz}(z)
\label{eq:TS}
\end{equation}
as a boundary mismatch, so that the boundary conditions read
\begin{align}
H_m'(0)+i\frac{\epsilon_{\mathrm{eff}}q_{s,m}}{\epsilon_s}H_m(0)&=-S_m(0),
\nonumber\\
H_m'(t)-i\frac{\epsilon_{\mathrm{eff}}q_{a,m}}{\epsilon_a}H_m(t)&=-S_m(t).
\label{eq:Tbc}
\end{align}
These conditions hold as written for $m\neq0$. For $m=0$ they apply to the modulation-induced part $H_0-H_0^{\mathrm{bg}}$, where the background field $H_0^{\mathrm{bg}}$ of the uniform layer carries the incident wave.
Green's identity $\int_0^t(v\mathcal{L}H-H\mathcal{L}v)\,dz=[vH'-v'H]_0^t$ together with Eq.~(\ref{eq:Tbc}) gives the diffracted amplitudes
\begin{align}
r_m&=\frac{1}{W_m}\left[\int_{0}^{t}v_m(z)F_m(z)\,dz+S_m(t)-v_m(0)S_m(0)\right],
\label{eq:Trm}\\
t_m&=\frac{1}{W_m}\left[\int_{0}^{t}u_m(z)F_m(z)\,dz+u_m(t)S_m(t)-S_m(0)\right].
\label{eq:Ttm}
\end{align}
The volume integrals correspond to the Green-function projections of Eqs.~(\ref{eq:rm}) and (\ref{eq:tm}). The interface terms have no TE counterpart. For TE the matching condition is diagonal in $m$, $S_m\equiv0$, and Eqs.~(\ref{eq:Trm}) and (\ref{eq:Ttm}) reduce to the volume integrals.

\subsection{Direct and indirect contributions}

Both $F_m$ and $S_m$ are sums over the harmonic index $n$. In both, the term $n=0$ couples the zeroth harmonic directly to the $m$th harmonic through the single Fourier component $\eta_m$,
\begin{equation}
F_m^{\mathrm{dir}}(z)=-\frac{\eta_m}{\eta_0}H_0''(z),
\qquad
S_m^{\mathrm{dir}}(z)=\frac{\eta_m}{\eta_0}H_0'(z).
\label{eq:Tdir}
\end{equation}
The remaining terms define the indirect contributions $F_m^{\mathrm{ind}}$ and $S_m^{\mathrm{ind}}$. Because Eqs.~(\ref{eq:Trm}) and (\ref{eq:Ttm}) are linear in $F_m$ and $S_m$, the separation carries over to the amplitudes,
\begin{equation}
r_m=r_m^{\mathrm{dir}}+r_m^{\mathrm{ind}},
\qquad
t_m=t_m^{\mathrm{dir}}+t_m^{\mathrm{ind}},
\label{eq:Tsplit}
\end{equation}
each part being obtained by inserting the corresponding pieces of $F_m$ and $S_m$. The direct contribution involves $H_0''$ in the volume source and $H_0'$ in the interface term. For $m\neq0$, the extinction of the $m$th channel requires the same condition as for TE,
\begin{equation}
\left|r_m^{\mathrm{dir}}\right|=\left|r_m^{\mathrm{ind}}\right|,
\qquad
\arg\left(r_m^{\mathrm{dir}}\right)-\arg\left(r_m^{\mathrm{ind}}\right)=\pi .
\label{eq:Tcancel}
\end{equation}

For $m=0$ the term $n=0$ is absorbed into the background, and the zeroth-order channel carries no direct contribution. The total zeroth-order amplitudes are $r_0^{\mathrm{bg}}+r_0$ and $t_0^{\mathrm{bg}}+t_0$, where the background parts are those of a uniform layer of dielectric constant $\epsilon_{\mathrm{eff}}$ between the substrate and the air. With the incident magnetic-field amplitude normalized to unity, the diffraction efficiencies are $R_m=(q_{s,m}/q_{s,0})|r_m|^2$ and $T_m=(\epsilon_sq_{a,m})/(\epsilon_aq_{s,0})\,|t_m|^2$, where these total amplitudes are used for $m=0$.

\subsection{Numerical verification}

The decomposition is evaluated using the same truncated field harmonics as the RCWA calculation. At the TM solution analyzed in Sec.~\ref{sec:tm}, with $\epsilon_h=\epsilon_s=12.3904$, $\rho=0.5703$, $t/\Lambda=0.6813$, and $\Omega=0.4295$, Eq.~(\ref{eq:Trm}) reproduces the RCWA amplitude $r_1$ to within $3\times10^{-12}$ at $\Omega=0.4295$ and $0.4009$ with $M=32$ harmonics on each side. The zeroth-order background is the uniform layer with $\epsilon_{\mathrm{eff}}=2.102$.
\section{Numerical search for the perfect-reflection solutions}
\setcounter{equation}{0}
\label{sm:tworidge}

\begin{table}[t]
\caption{Perfect-reflection solutions of the two-ridge grating in the four-channel window, calculated by RCWA with $M=32$ for $\epsilon_h=\epsilon_s=12.3904$.}
\label{tab:S1}
\centering
\begin{tabular*}{\columnwidth}{@{\extracolsep{\fill}}c@{\hspace{2\tabcolsep}\hspace{-1.5em}}cccc}
\hline
Solution & $\rho_1$ & $\rho_2$ & $t/\Lambda$ & $\Omega$ \\
\hline
1 & 0.4537 & 0.0677 & 0.4328 & 0.9748 \\
2 & 0.0117 & 0.4049 & 0.6022 & 0.9443 \\
3 & 0.0308 & 0.6083 & 0.7776 & 0.8977 \\
4 & 0.4028 & 0.1379 & 0.9485 & 0.9232 \\
5 & 0.5349 & 0.0092 & 0.9573 & 0.8917 \\
6 & 0.0353 & 0.3997 & 0.9850 & 0.8927 \\
7 & 0.1440 & 0.4030 & 1.0208 & 0.9968 \\
8 & 0.5425 & 0.0148 & 1.0410 & 0.9556 \\
9 & 0.0532 & 0.4348 & 1.1668 & 0.8760 \\
10 & 0.1642 & 0.0978 & 1.1908 & 0.8715 \\
11 & 0.0587 & 0.3503 & 1.2117 & 0.8588 \\
12 & 0.0652 & 0.4448 & 1.2704 & 0.9171 \\
13 & 0.0995 & 0.1565 & 1.3092 & 0.9237 \\
14 & 0.3909 & 0.1721 & 1.3815 & 0.8572 \\
15 & 0.1004 & 0.1503 & 1.4230 & 0.9698 \\
\hline
\end{tabular*}
\end{table}

The TE perfect-reflection solutions for $\epsilon_h=\epsilon_s=12.3904$ were obtained by solving the cancellation conditions from regular grids of initial points. Each search minimized a residual vector built from the power-normalized outgoing amplitudes in the open channels other than the incident one, whose squared norm equals $1-R_0$ for a lossless grating, using a Levenberg--Marquardt iteration with a finite-difference Jacobian. Only converged points with $1-R_0<10^{-12}$ were retained, and coincident solutions were counted once. For the single-ridge grating in the three-channel window, $18\,432$ initial points spanning $0.05\leq\rho\leq0.95$ and $0.05\leq t/\Lambda\leq2.50$ yielded $25$ distinct solutions, with $\rho$ between $0.06$ and $0.66$ and $t/\Lambda$ between $0.20$ and $3.43$. The thinnest of these solutions is the TE design of Refs.~\cite{Gebski2015,Dems2017}, whose published parameters, used in Figs.~\ref{fig1}--\ref{fig3}, agree with it to the precision given in Refs.~\cite{Gebski2015,Dems2017}. Because the search starts from a finite set of initial points, further solutions, particularly at larger $t/\Lambda$, cannot be excluded. In the two-channel window, the five branches, with $\rho$ between $0.45$ and $0.52$ and $t/\Lambda$ between $0.93$ and $3.10$, were traced by arc-length continuation, and perfect reflection holds to numerical precision along all of them. In the four-channel window, neither a grid search over $1.9\times10^6$ points with $0.10\leq\rho\leq0.90$ and $0.05\leq t/\Lambda\leq1.50$ nor a search from $4\,608$ initial points with $0.05\leq\rho\leq0.95$ and $0.02\leq t/\Lambda\leq1.00$ found a solution.

Table~\ref{tab:S1} lists the $15$ TE perfect-reflection solutions of the two-ridge grating found in the four-channel window $3/\sqrt{\epsilon_s}<\Omega<1$ for $\epsilon_h=\epsilon_s=12.3904$. The ridges of widths $\rho_1\Lambda$ and $\rho_2\Lambda$ are centered at $x=0$ and on the unit-cell boundary $x=\pm\Lambda/2$. The search started from $9\,888$ initial points of $(\rho_1,\rho_2,t/\Lambda,\Omega)$ with $0.05\leq\rho_1\leq0.90$, $0.02\leq\rho_2\leq0.60$, $\rho_1+\rho_2\leq0.98$, and $0.05\leq t/\Lambda\leq1.20$, and the solutions are listed in order of increasing $t/\Lambda$. The Jacobians of the conditions have ranks two, three, and four at the solutions in the two- and three-channel windows and for the two-ridge grating, respectively, and these ranks are unchanged for truncation orders up to $|m|\leq128$. The coordinates are rounded to four decimal places. Before rounding, $1-R_0$ of each solution is below $3\times10^{-14}$ with $M=32$.

\begin{table}[t]
\caption{TM perfect-reflection solutions of the single-ridge grating in the three-channel window, calculated by RCWA with $M=32$ for $\epsilon_h=\epsilon_s=12.3904$.}
\label{tab:TM}
\centering
\begin{tabular*}{\columnwidth}{@{\extracolsep{\fill}}c@{\hspace{2\tabcolsep}\hspace{-1.5em}}ccc}
\hline
Solution & $\rho$ & $t/\Lambda$ & $\Omega$ \\
\hline
1 & 0.3168 & 0.6478 & 0.5738 \\
2 & 0.5791 & 1.1354 & 0.7633 \\
3 & 0.5968 & 1.1581 & 0.7786 \\
4 & 0.6222 & 1.3189 & 0.7617 \\
5 & 0.2795 & 1.7210 & 0.6141 \\
6 & 0.5826 & 1.7807 & 0.7607 \\
7 & 0.6217 & 2.0213 & 0.7209 \\
8 & 0.6590 & 2.2593 & 0.7306 \\
9 & 0.6768 & 2.3177 & 0.6880 \\
10 & 0.6698 & 2.4021 & 0.7239 \\
\hline
\end{tabular*}
\end{table}

The TM solutions for the same dielectric constants were obtained in the same way. In the three-channel window, $2\,800$ initial points spanning $0.04\leq\rho\leq0.96$ and $0.05\leq t/\Lambda\leq2.50$ yielded the ten solutions listed in Table~\ref{tab:TM}, and, as in the TE search, further solutions cannot be excluded. In the two-channel window, the three branches of Fig.~\ref{fig6}(d) were traced by arc-length continuation, and perfect reflection holds to numerical precision along all of them. The coordinates in Table~\ref{tab:TM} are rounded to four decimal places. Before rounding, $1-R_0$ of each solution is below $5\times10^{-13}$ with $M=32$.

\end{document}